\documentclass[pre,notitlepage, twocolumn]{revtex4-1}
\usepackage{amsmath}
\usepackage{amssymb}
\usepackage{graphicx}
\usepackage{enumerate}
\usepackage{color}
\usepackage[colorlinks = true,
            linkcolor = blue,
            urlcolor  = blue,
            citecolor = blue,
            anchorcolor = blue]{hyperref}
\usepackage{physics}
\usepackage{placeins}
\usepackage{booktabs, tabularx}
\usepackage{subcaption}

\def\doi#1#2{\href{https://doi.org/#2}{#1}}
\def\hri#1#2{\href{http://arxiv.org/abs/#1}{[arXiv:#1]}}

\newcommand{\eq}[1]{\begin{equation} #1 \end{equation}}

\newcolumntype{Y}{>{\centering\arraybackslash}X}

\AtBeginDocument{
\heavyrulewidth=.08em
\lightrulewidth=.05em
\cmidrulewidth=.03em
\belowrulesep=.65ex
\belowbottomsep=0pt
\aboverulesep=.4ex
\abovetopsep=0pt
\cmidrulesep=\doublerulesep
\cmidrulekern=.5em
\defaultaddspace=.5em
}
\begin{document}

\title{Application of machine learning in Bose-Einstein condensation critical-temperature analyses of path-integral Monte Carlo simulations}
\author{Adith Ramamurti}
\email{adith.ramamurti@alumni.stonybrook.edu}
\affiliation{Department of Physics and Astronomy \\ Stony Brook University \\ Stony Brook, NY 11794-3800}
\altaffiliation[Present address: ]{U.S. Naval Research Laboratory, Washington, D.C. 20375}
\date{December 18, 2019}
\begin{abstract}
We detail the use of simple machine learning algorithms to determine the critical Bose-Einstein condensation (BEC) critical temperature $T_\text{c}$ from ensembles of paths created by path-integral Monte Carlo (PIMC) simulations. We quickly overview critical temperature analysis methods from literature, and then compare the results of simple machine learning algorithm analyses with these prior-published methods for one-component Coulomb Bose gases and liquid $^4$He, showing good agreement.
\end{abstract}

\maketitle

\section{Introduction}

The path-integral description of quantum mechanics was first developed by Feynman, describing the probability amplitude for a system to go from an initial-time state to a final-time state as a sum over all possible trajectories from the initial state to the final state weighted by their corresponding action \cite{Feynman:1948ur}.
In terms of a system constituted of particles, the action includes the trajectories of the particles themselves, known as the kinetic action, as well as the equal-time interaction between particles, known as the potential action. 

This description was subsequently extended to equilibrium finite-temperature systems.
At finite-temperature, a system can be thought to be in the space $R^3\times S^1$, where the Euclidean (imaginary) time dimension $\tau$ is periodic with length $\beta = 1/T$, the inverse temperature. 
This periodic time direction is sometimes known as the Matsubara circle.
Feynman showed that, by calculating periodic path integrals for particles in Euclidean time, one can find the density matrix, and by extension the thermodynamics, of the system. 

An early success of the finite-temperature path integral was to the phenomenon of Bose-Einstein condensation (BEC) in liquid helium \cite{PhysRev.91.1291,Feynman:1953zz}, which is expressed physically as the $^4$He atoms being in a superfluid  state (long-range delocalization). 
At low temperatures, the paths are long, and a particle path can deviate further from the classical (straight) trajectory with less penalty in the action, compared to the same particle at high-temperatures.
Feynman drew a connection between the appearance of ``ring polymers'' or ``Bose clusters,'' permutation chains of the paths of bosons, and the appearance of a condensate (atoms in the superfluid state) at near-zero temperatures.
When $k$ bosonic paths are permuted, the particle path wraps $k$ times around the Matsubara circle, each time ending up at the location of another boson at $\tau = 0$, before returning to its original location.
During such a trajectory, the position of the path in space can deviate far from its original location, which is the manifestation of the delocalization/superfluid phenomenon.

For a general interacting system, however, evaluation of the path integral is non-trivial, as the actions become complicated. 
As such, numerical techniques emerged as the primary avenue to tackle the problem. 
Numerical methods for quantum mechanics were pioneered by Creutz and Freedman \cite{Creutz:1980gp}, and in the QCD community, for example, evaluation of finite-temperature path integrals on the lattice remains a fundamental method for understanding nuclear phenomena.

Path-integral Monte Carlo (PIMC) is another numerical technique for evaluating path integrals for many-body systems at finite temperature, with widespread use in the condensed matter and chemistry communities. 
Unlike lattice simulations, where all four coordinates $(\tau,x,y,z)$ are discretized, PIMC simulations are continuous (and periodic) in the spatial dimensions and only discretized in the imaginary-time direction.
A full description of path-integrals and PIMC is outside of the scope of this work; a thorough review of the PIMC numerical technique and the theory behind it, as applied to liquid-$^4$He, can be found in Ref. \cite{RevModPhys.67.279}. 

\section{Benchmark analyses}

For BEC critical temperature analyses, the classic technique used in conjunction with PIMC is the finite-size scaling of the superfluid fraction, first described by Ceperley, Pollock, and Runge \cite{PhysRevB.36.8343,PhysRevB.46.3535}. 
In this method, the fraction of particles that are in a superfluid state is calculated for a range of temperatures around the critical temperature.
The superfluid fraction at a particular temperature, found from an ensemble of path configurations, is given by
\eq{
\frac{\rho_\text{s}}{\rho} = \frac{m}{\hbar} \frac{\langle \vec W^2\rangle \cdot \vec L^2}{3 \beta N}\,,
}
where $N$ is the number of particles, $m$ is the mass of the particles, $\vec L$ the displacement vectors from a point in the simulation box to its periodic images, $\vec W$ the total winding number of paths in a particular configuration through/around the periodic simulation box, and $\langle\ldots\rangle$ indicating an average over all path configurations. 

To clarify this for the reader, if a particle trajectory at $\tau=0$ is located at $(x,y,z)$ and ends at its periodic image $(x+nL_x,y,z)$ at $\tau = \beta$, it has a winding number of $n \in (\ldots,-1,0,1,\ldots)$ in the $x$ direction and 0 in $y,z$.
$\vec W$ is then the sum of the winding numbers of all paths in a configuration; if one path has winding 1 in the $x$ direction, and another has $-1$ in the $x$ direction, the total $W_x = 0$.
One can either pick a direction, say $x$, to compute this quantity $\langle \vec W^2 \rangle \cdot \vec L^2 \equiv \sum_n \langle W_{x,n}^2 \rangle L_{x,n}^2$; or compute an average over all directions $\langle \vec W^2 \rangle \cdot \vec L^2\equiv (\sum_{i,n} \langle W_{i,n}^2 \rangle L_{i,n}^2)/3$.

The ``amount'' of long-range delocalization of a path can be measured by the number of times the path winds around the periodic simulation box, and thus is a proxy for the fraction of particles in the superfluid state (i.e. in the condensate).
In an infinite system, one expects that, below the critical temperature, there will be an infinite-length permutation cycle allowing a non-zero winding number, while above $T_\text{c}$, there is no such possibility. 
For a finite-but-periodic system, as is studied in numerical simulations, the hard cut-off is smeared out as particles only need to form a ``long-enough'' permutation chain to wind around the finite periodic box.
Nevertheless, we can use finite-size scaling arguments to find the infinite-size limit for the transition temperature.

The superfluid fraction for particles in a periodic box of side-length $L$ will scale as
\eq{
\frac{\rho_\text{s}}{\rho}(T,L) = L^{-1}Q(L^{-1/\nu}t)\,,
}
for some function $Q$ and $t = (T-T_\text{c})/T_\text{c}$.
Therefore, all functions of the form
$L \frac{\rho_\text{s}}{\rho}(T,L)$
 should cross at $T = T_\text{c}$. 
Minor corrections to this scaling are given in Ref. \cite{PhysRevB.46.3535}.

Another approach, formulated by D'Alessandro, D'Elia, and Shuryak for lattice simulations \cite{D'Alessandro:2010xg}, related the scaling of permutation-cycle statistics (i.e. the probability of finding a path permuted with $k-1$ other paths in any given configuration state) to the critical temperature. 
The density of $k$-cycles $\rho_k$ is defined in terms of the probability of finding a particle in a $k$-cycle, and is given by
\eq{
\rho_k(T) =  \frac{N}{kV} P_k(T) = f(k) e^{- k \hat{\mu}} \,, \label{eq:muhat}
}
where $N$ is the number of particles, $V$ is the volume of the simulation box, $\hat\mu \equiv \mu/T$ is an effective chemical potential scaled by temperature, and $f$ some decreasing function of $k$ with form $f(k) \sim 1/k^\gamma$.
The critical temperature is defined as where $\mu$ vanishes, indicating that the exponential suppression of the permutation cycles no longer exists, and it is the least ``costly'' in terms of the action for particles to permute.
The $\hat\mu$s can be fit as a function of the temperature, 
\eq{
\hat\mu(T) = A(T-T_\text{c})^\nu\,.
}

It was shown in Ref. \cite{Ramamurti:2017fdn}  -- see therein for a slightly more thorough overview of these analyses, as well -- that both of these methods yield the same critical temperature for PIMC simulations of one- and two-component quantum Coulomb Bose gases, as well as for liquid $^4$He.

We seek to compare simple machine learning approaches with the analysis methods outlined above. 
The superfluid fraction method requires at least three system sizes run at many temperatures below and above $T_\text{c}$ to accurately fit the superfluid fraction as a function of $T$ and study the finite-size scaling. 
The permutation-cycle method, while requiring runs only above the critical temperature and of only one system size, requires parameter-tuning necessary to get a good fit for the data, in terms of choosing which $k$-cycles to use in the fits to extract $\hat\mu$ at each temperature and which temperature ranges to use when fitting $\hat\mu$. 
Machine learning algorithms, when given a good training dataset, can analyze large amounts of data with minimal user input, and thus allow, possibly, for a more consistent method of analysis.

We will utilize the data from the simulations detailed in Ref. \cite{Ramamurti:2017fdn} in our comparison. 
In particular, we will study a one-component quantum Coulomb Bose gas, with particles of charge 1, inter-particle spacing 1  -- which defines the box size as a function of number of particles --, mass 1, and with $\hbar = 1$ and $k_B = 1$. 
In these units, the critical condensation temperature for a non-interacting gas of bosons is 
\eq{
T_\text{c} =2\pi \left(\frac{1}{\zeta(3/2)}\right)^{2/3}  = 3.3125\,.
}

Before getting to the machine learning analysis, we will give a quick example of the established methods. Let us focus on the case where the interaction between the particles is given by $\alpha/r^2$ with $\alpha = 2$. Using 30,000 independent path configurations using PIMC at each temperature in the range $1.6\ldots5.1$, we can extract average winding number and permutation statistics for each temperature.

\begin{figure}[h!]
\includegraphics[width=.48\textwidth]{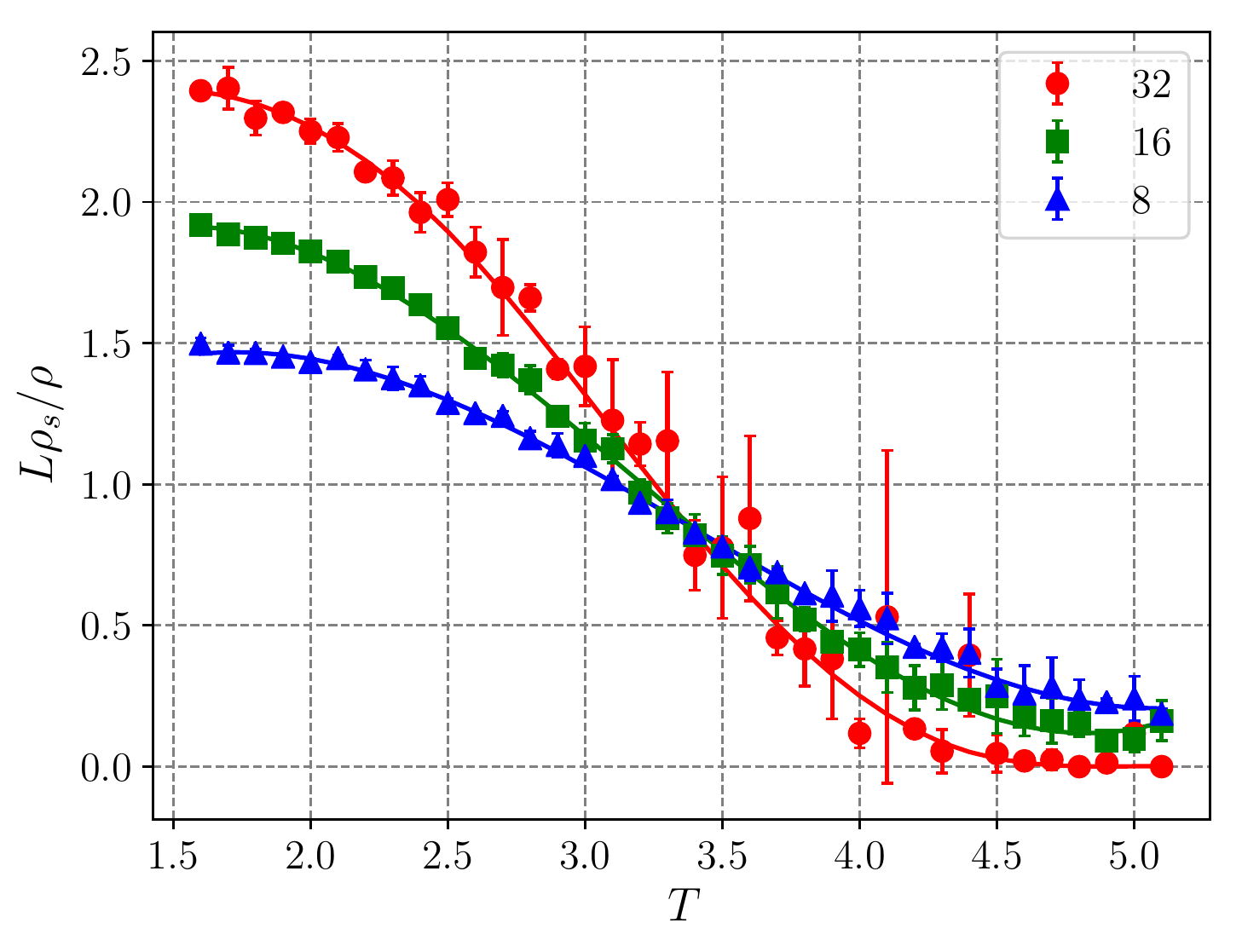}
\caption{Scaled superfluid fraction calculated for various temperatures for system sizes of 8 (blue triangles), 16 (green squares), and 32 (red circles) quantum Coulomb particles with coupling $\alpha = 2$, along with curve fits.}
\label{fig:fss}
\end{figure}

Fig. \ref{fig:fss} shows the scaled superfluid fraction of three different system sizes of Coulomb-interacting particles with coupling $\alpha = 2$ as a function of temperature.
 As stated earlier, the critical temperature is found where the scaled superfluid fraction curves intersect for systems of different number of particles. 
 In this case, the temperature is found to be $3.42\pm0.025$, slightly higher than the non-interacting case; this deviation is expected \cite{PhysRevLett.79.3549}.

\begin{figure}[h!]
 \begin{subfigure}[b]{0.48\textwidth}
 \centering
\includegraphics[width=\textwidth]{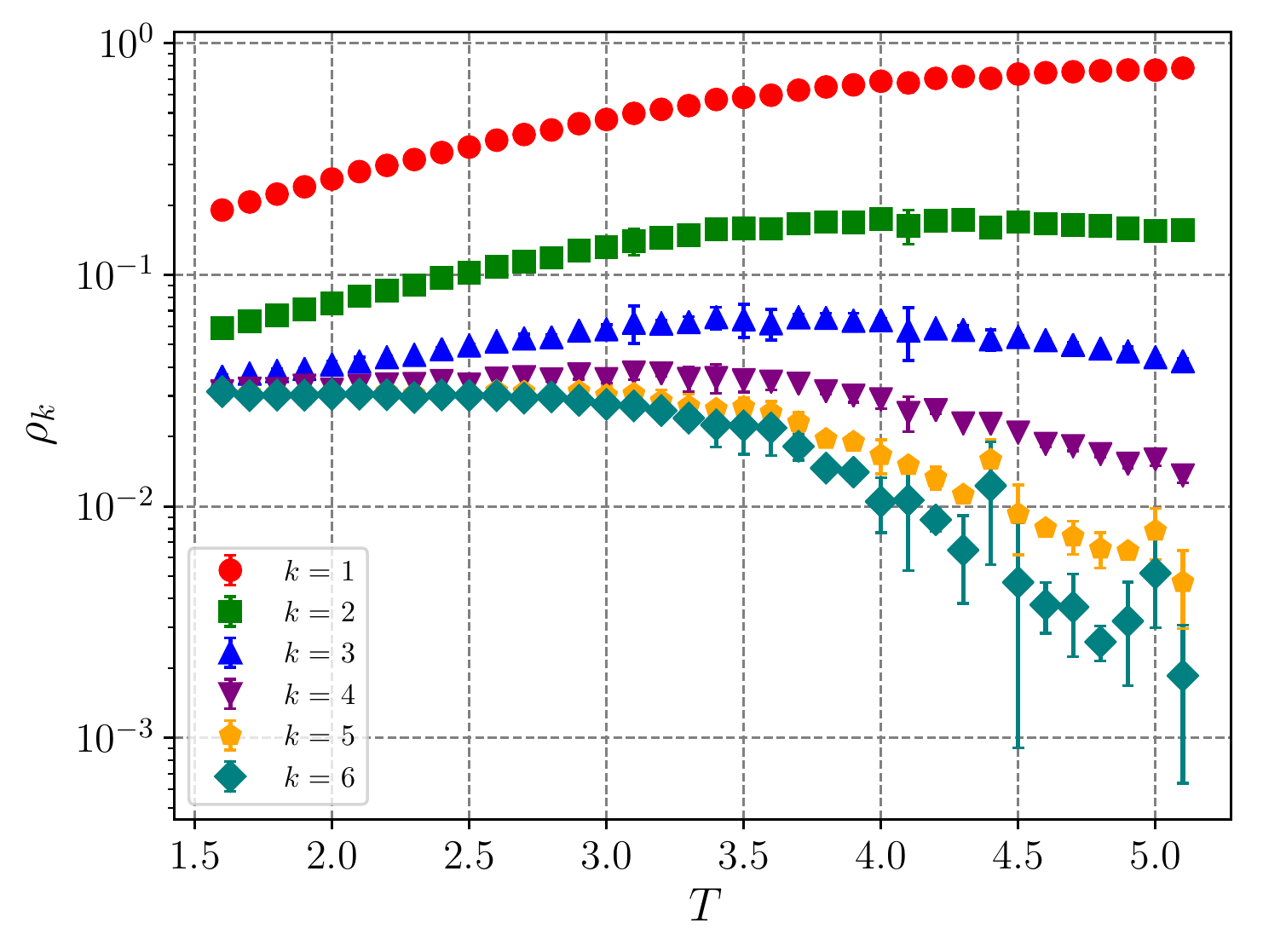}
\caption{}
\label{fig:pcs1}
\end{subfigure}
 \begin{subfigure}[b]{0.48\textwidth}
 \centering
\includegraphics[width=\textwidth]{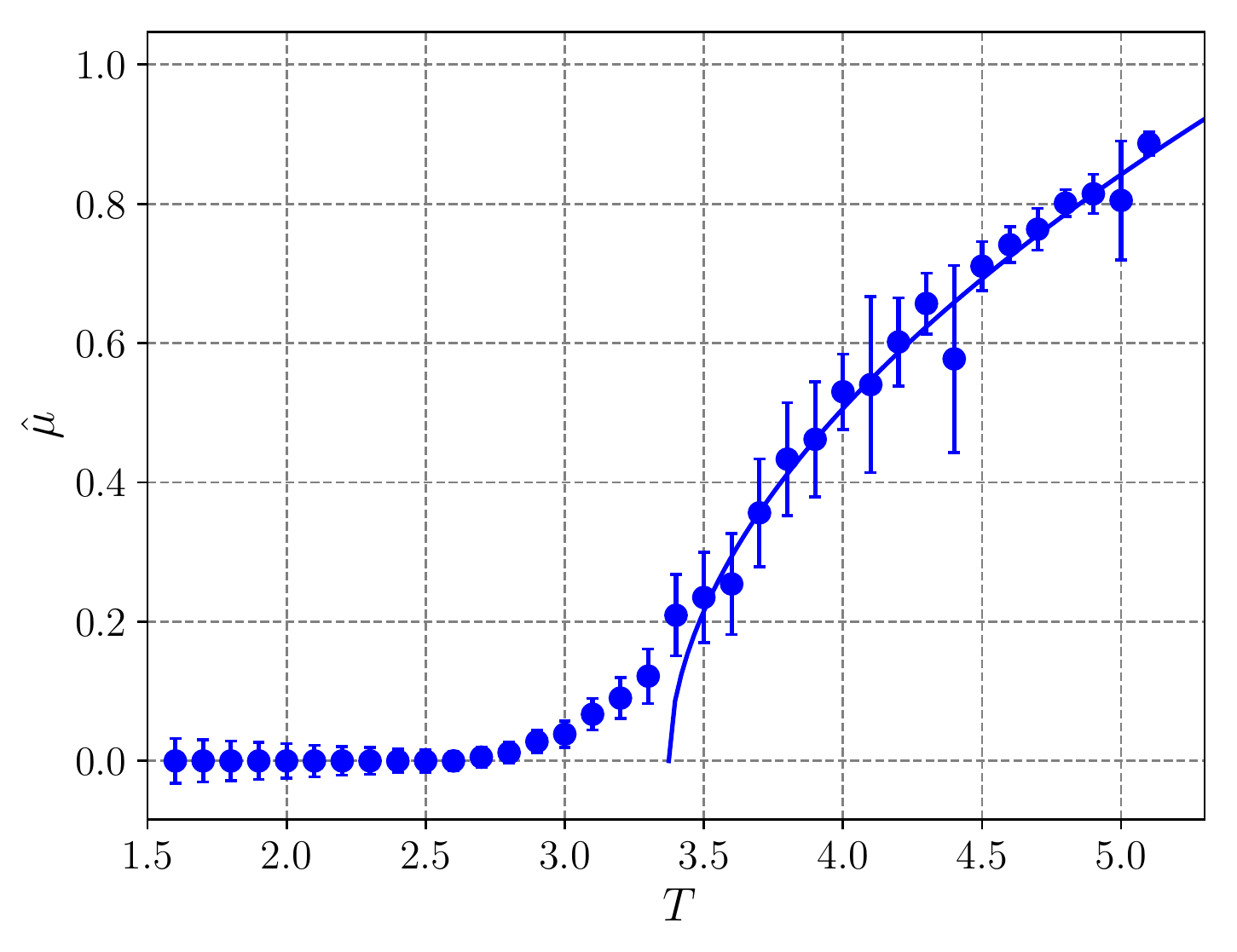}
\caption{}
\label{fig:pcs2}
\end{subfigure}
\caption{(a) Density of $k$-cycles for $k=1\ldots6$ as a function of temperature, and (b) the fitted $\hat\mu$ from the $k$-cycle densities and subsequent fit to find the critical temperature.}
\end{figure}

Fig. \ref{fig:pcs1} shows the density of $k$-cycles for $k=1\ldots6$ for the same system. As an example of the disappearing suppression of permutations, as temperature decreases, the density of $k=6$ cycles levels off at around 3.3--3.5 and remains roughly constant down to low temperatures. Ideally, without numerical effects, all $k$-cycles would have a probability $1/N$ at zero temperature for a system of $N$ particles. From these densities, at each temperature, one can find the scaled effective chemical potential $\hat{\mu}$ by fitting with Eq. \ref{eq:muhat} with $\gamma = 2$. The critical temperature, defined as where $\hat\mu$ vanishes, is found from fitting the above-$T_\text{c}$ portion of the $\hat\mu$s, shown in Fig. \ref{fig:pcs2}. This yields a critical temperature of $3.374\pm0.06$, almost identical to the prior method.

See Ref. \cite{Ramamurti:2017fdn} for more discussion of the one-component Coulomb Bose gas, as well as application of these methods to a two-component Coulomb Bose gas and liquid-$^4$He.

\section{Machine learning analyses}

Finally, we can use machine learning on the same data. 
For this paper, we only use support-vector machine (SVM) classifiers   \cite{berwick2003idiot} and do a simple optimization over model parameters.
SVM classifiers essentially project an $n$-dimensional dataset onto an $n+m$-dimensional space, and finds a best-fit hyperplane that divides the data into two classifications.
This choice of model and the parameters used can be tweaked in practice; we are simply trying to illustrate applicability of these approaches to this problem.

We train the classifier (with classifications of above and below $T_\text{c}$) on the non-interacting case, for which we know the analytic answer, to look for patterns in permutation cycles and winding number in condensed (below-$T_\text{c}$) and non-condensed states (above-$T_\text{c}$).
Specifically, we can pass in ensemble-averaged statistics using a jackknife-like sampling method: we take our 30,000 ensembles for each temperature and randomly organize them in groups of 200 and generate averaged statistics.
We will denote these ensemble-averaged statistics as ``averaged configurations'' in what is to follow.

\begin{figure*}[htbp]
\includegraphics[width=\textwidth]{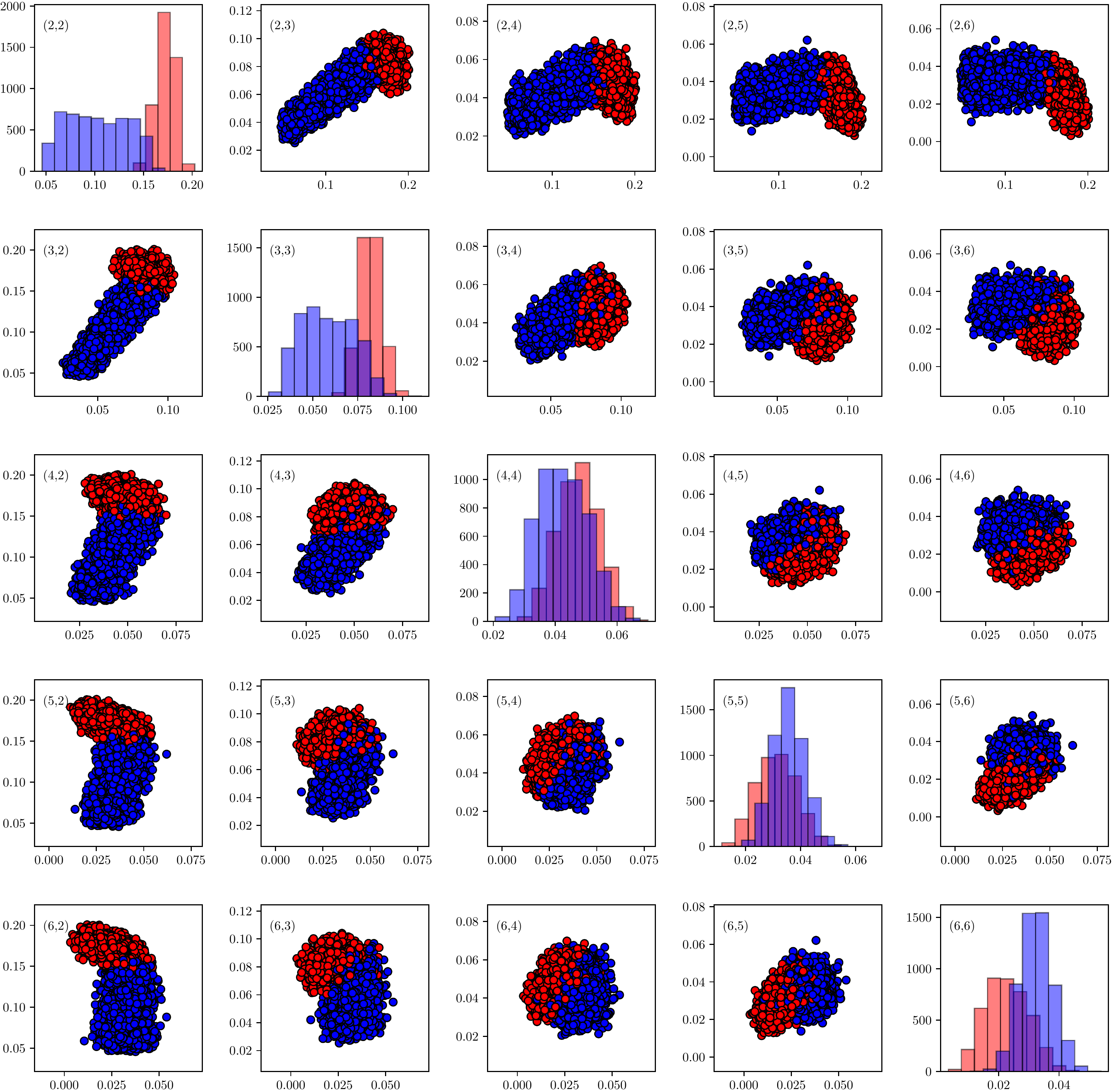}
\caption{Visualization of ensemble averaged probabilities of various pairs of $k$-cycles. The ensemble averages for systems below $T_\text{c}$ are denoted by blue dots, while averages for those above are denoted by red dots.}
\label{fig:mlsep}
\end{figure*}

\begin{figure*}[htpb]
\includegraphics[width=\textwidth]{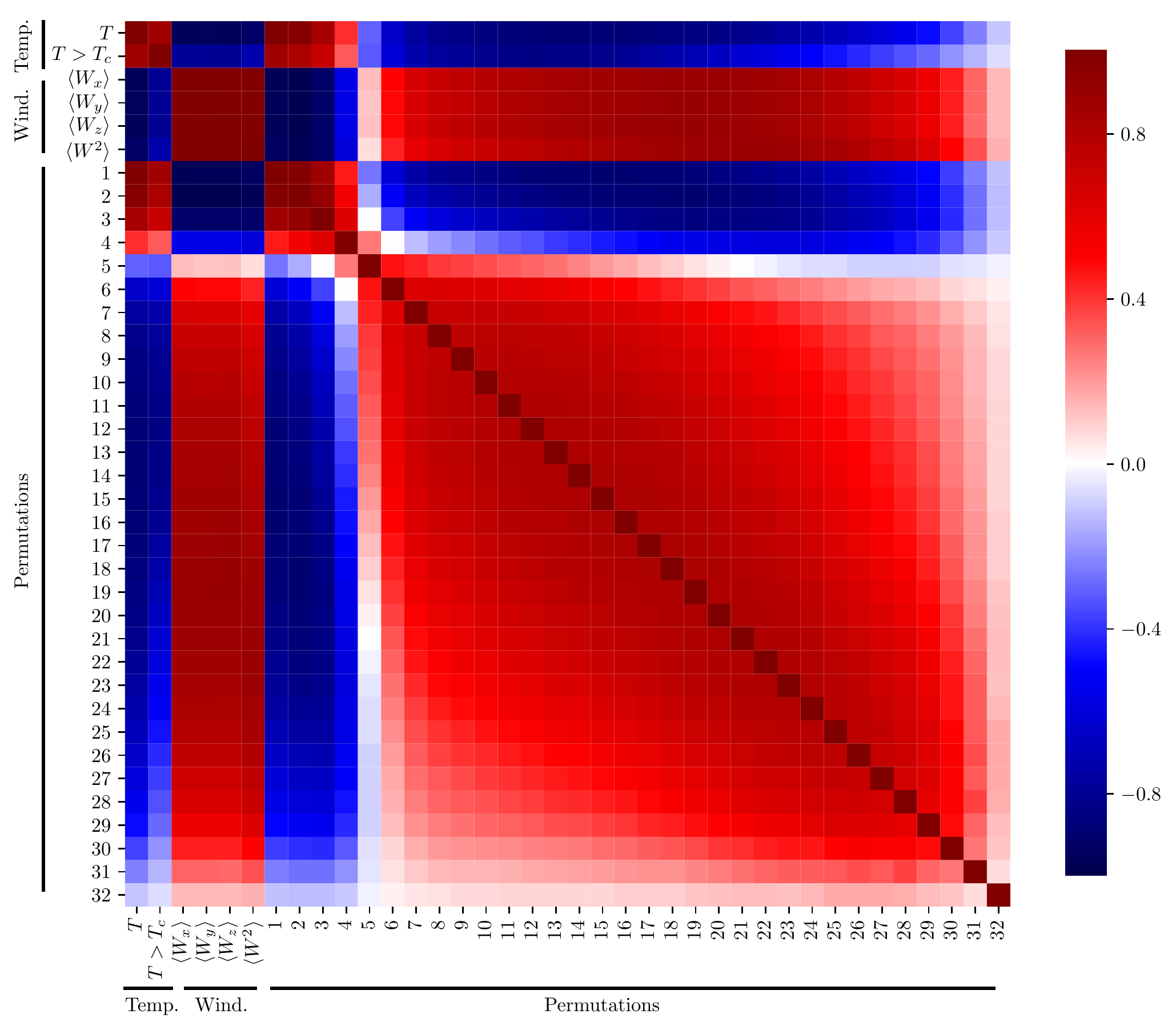}
\caption{Correlations between temperature, $T>T_\text{c}$, average winding numbers, and permutation cycle probabilities from 1-32 (in that order left-to-right and top-to-bottom).}
\label{fig:mlcorr}
\end{figure*}

We now show two visualizations of the non-interacting 32-particle simulation data we are using as a training data set. 
Fig. \ref{fig:mlsep} shows the distribution of below $T_\text{c}$ (blue) and above $T_\text{c}$ (red) ensemble averaged data projected onto the space of permutation-cycle probablilites.
The plot labeled (2,3), for example, has the probability of a particle in a configuration being in a 2-cycle on the $x$-axis and the probability of a particle in a configuration being in a 3-cycle on the $y$-axis.
Each averaged configuration is plotted as a function of these two probabilities.
We can see that some pairs of permutation probabilities display a sharp separation in the critical temperature classification; in these, a simple classification scheme can be based on a line in that particular subspace. 
Other pairs of cycle probabilities have a more complex relationship.
 
Fig. \ref{fig:mlcorr} shows correlations between temperature, winding, and permutation cycles. 
A few observations for this particular system of 32 particles: 
\begin{enumerate}
\item Temperature is highly correlated with the presence of 1- and 2-cycles, and anti-correlated with cycles $k>5$. 
\item Temperature is heavily anti-correlated with winding. 
\item Statistics of permutation cycles of a certain size are correlated with those of other permutation cycles with sizes similar (i.e. probabilities of 9-cycles and 10-cycles are correlated). 
\item Finite-size effects start playing a role for cycles $k\gtrsim18$. Without these effects, for temperatures under the critical temperature, we should see $f(k) \sim 1/k^\gamma$ behavior out to $k = \infty$. The finite-size effects result in, for example, temperature having little-to-no correlation with $k=31,32$ cycles, when we know, in reality, that it should be highly anti-correlated with all large $k$.
\end{enumerate}

The first two points directly correspond to the physics behind the two traditional methods of analysis.
This kind of visualization can also aid with the standard permutation cycle method, as it can indicate what range of $k$ to include in the fits.

We note that these observations specifically pertain to systems of 32 particles and the resulting machine learning model can only be applied to a system of the same size; when analyzing a system of $N$ particles, one needs a baseline of $N$ non-interacting particles.

We can now apply this model to the data analyzed above with the published methods. 
To do so, we take all of the averaged configurations at a particular temperature and allow the algorithm to classify each of them as above or below $T_\text{c}$.
An averaged configuration deemed to be below $T_\text{c}$ is assigned a 0 and one that is above a 1.
Then, we average the classification results for that temperature to create an overall ``probability" that the temperature is above the critical temperature.
To make this explicit, if the classifier says that 1/200 of the averaged configurations at a particular temperature are below $T_\text{c}$, that temperature is assigned 0.995, while if the classifier deems that 50/200  averaged configurations at a temperature  are below $T_\text{c}$, then that temperature is assigned a 0.25. 
We then fit this probability data as a function of temperature using the sigmoid function
\eq{
P(t) = S(A t ) \equiv \frac{e^{At}}{e^{At} + 1}\,,
} with $t = T-T_\text{c}$, to extract the critical temperature.
Using this method on the coupling $\alpha=2$ data studied above, the SVM classifier yields a result of $3.383\pm0.03$, which is within errors of the finite-size-scaling and permutation-cycle methods.

In Table \ref{tab:mlres}, we summarize results for the one-component quantum Coulomb Bose gas with a variety of couplings. We see excellent agreement between the established methods and the machine learning model for all coupling strengths studied.

\begin{table}
\centering
\begin{tabularx}{.5\textwidth}{X Y Y Y}
\toprule 
$\alpha$ & PC & SFF & SVM \\ \midrule
0 & 3.28 & 3.31 & -- \\ 
1 & 3.36 & 3.31 & 3.33 \\ 
2 & 3.37 & 3.42 & 3.38 \\ 
5 & 3.47 & 3.52 & 3.53 \\ 
10 & 3.55 & 3.63 & 3.61 \\ 
20 & 3.45 & 3.48 & 3.44 \\
50 & 3.31 & 3.31 & 3.28 \\ \bottomrule
\end{tabularx}
\caption{Critical temperature for a one-component Coulomb Bose gas with coupling $\alpha$ found using permutation cycle (PC), finite-size scaling of the superfluid fraction (SFF), and machine learning (SVM) methods. Errors for the first two methods are visible in Fig. 3 of Ref. \cite{Ramamurti:2017fdn}, and are $\pm0.03$ for all of the machine learning model results.}
\label{tab:mlres}
\end{table}

\begin{figure}[t!]
\includegraphics[width=.48\textwidth]{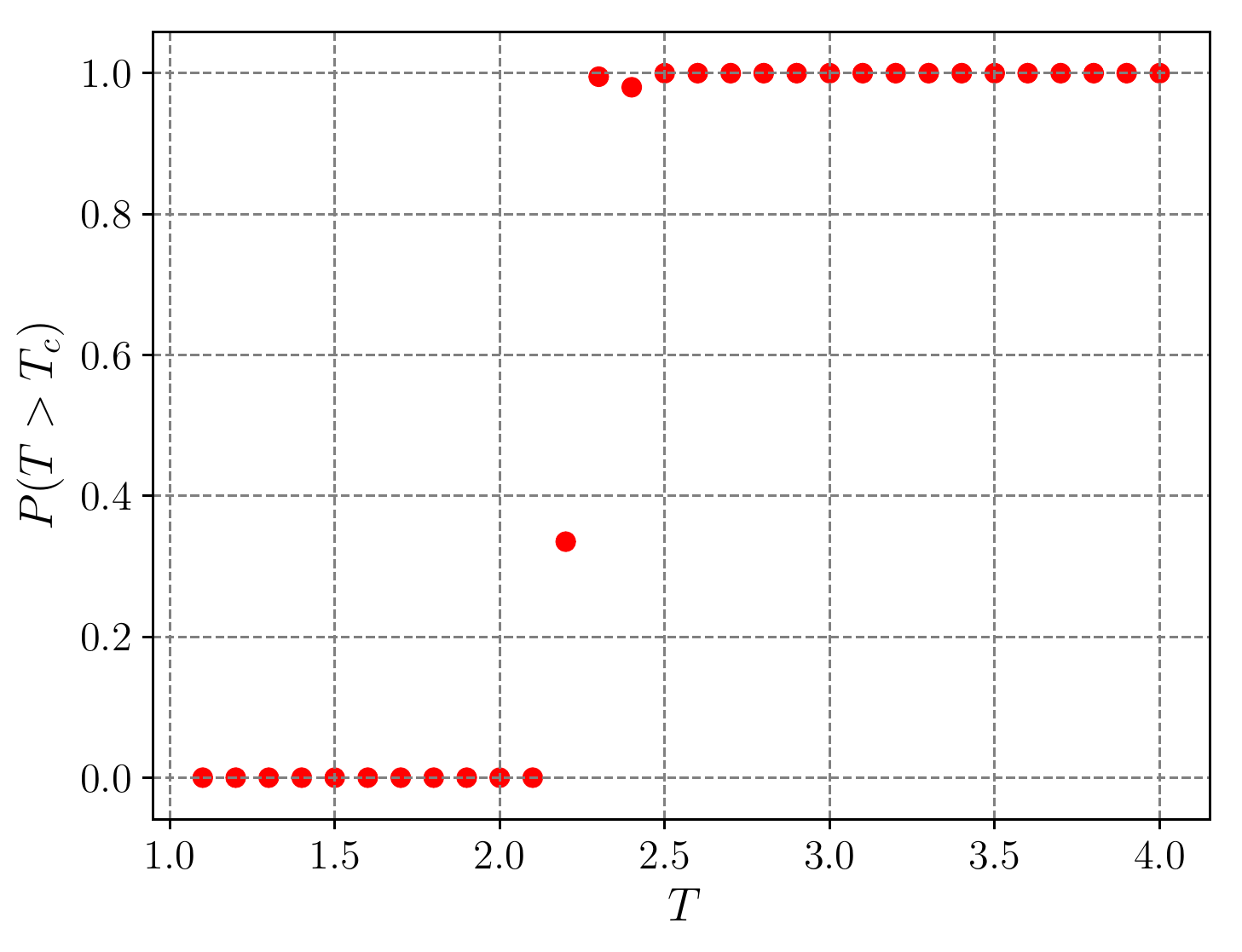}
\caption{Probability that a temperature is above the critical temperature for a system of 32 liquid helium particles, as classified by the machine learning algorithm.}
\label{fig:he4svm}
\end{figure}

As a final validation test of the method, we can use the machine learning algorithm to try to find the critical temperature for a system of $^4$He particles at SVP; the established experimental value of $T_\text{c}$ for this system is 2.17 K. PIMC simulations run with 64 helium particles and analyzed  with the superfluid fraction method give a critical temperature of $2.19\pm0.02$ K \cite{RevModPhys.67.279}, which is the most accurate result possible when using the empirical two-body Aziz potential, while the permutation cycle method coupled with simulations of 128 particles gives $2.21\pm0.04$ K \cite{Ramamurti:2017fdn}.

We use our pre-trained model, which only knows about the statistics of non-interacting bosons, on a system of 32 helium particles. The results of the classifier are shown in Fig. \ref{fig:he4svm}. The fit to the data gives a final result of $2.23\pm0.03$ K, which is within errors of the other methods.

\section{Summary}

We have shown how machine learning techniques can be used in conjunction with PIMC simulations to accurately find the Bose-Einstein condensation critical temperature of systems of interacting bosons. Machine learning techniques require far less input from the user, and can be quicker as well, as was the case for the analyses in this paper, while giving results that fall within statistical errors of well-established methods.

\vspace{1ex}

\textbf{Acknowledgements.} This analysis presented in this paper did not use U.S. Naval Research Laboratory resources or funding. The data analyzed comes from simulations carried out during the author's graduate studies, using the LI-red computational cluster at the Institute for Advanced Computational Science (IACS), Stony Brook University with support in part by the U.S. D.O.E. Office of Science, under Contract No. DE-FG-88ER40388. The author thanks Dr. Edward Shuryak for discussions of the simulation methods and resulting data.
 
The path-integral Monte Carlo code and analysis scripts for all three methods used for this paper are available at: \href{https://github.com/aramamurti/PIMC}{https://github.com/aramamurti/PIMC}.

\end{document}